\documentclass[aps,prd,superscriptaddress,nofootinbib,tighten,preprint]{revtex4}
\usepackage[utf8]{inputenc}
\usepackage[english]{babel}
\usepackage{amsmath}
\usepackage{subfigure}
\usepackage{cancel}
\usepackage{amsfonts}
\usepackage{amssymb}
\usepackage{graphicx}
\usepackage{hyperref}
\usepackage{lineno,color,slashed,changes,color}
\newcommand{\be}{\begin{equation}}
\newcommand{\ee}{\end{equation}}
\newcommand{\bea}{\begin{eqnarray}}
\newcommand{\eea}{\end{eqnarray}}

\catcode`@=12

\hypersetup{
	colorlinks = true,
	linkcolor = blue,
	citecolor = magenta
}

\begin{document}

\title{Higgs-portal vector dark matter at a low reheating temperature}

\author{Sarif Khan}
\email{sarifkhan@cau.ac.kr}
\affiliation{Department of Physics, Chung-Ang University, Seoul 06974, Korea.}
\author{Jongkuk Kim}
\email{jongkukkim@cau.ac.kr}
\affiliation{Department of Physics, Chung-Ang University, Seoul 06974, Korea.}
\author{Hyun Min Lee}
\email{hminlee@cau.ac.kr}
\affiliation{Department of Physics, Chung-Ang University, Seoul 06974, Korea.}

\begin{abstract} 

We study vector dark matter (DM) production with Higgs-portal type interactions 
in the scenarios with a low reheating temperature which can be realized by a prolonged decay of the inflaton after inflation.
We take the reheating temperature to be large enough to match the observations in Standard Cosmology such as Big Bang Nucleosynthesis but small enough below the DM mass for the DM production.
We analyze the impact of the model parameters including the extra gauge coupling and the reheating temperature on the DM relic density, collider bounds and DM direct and indirect detection experiments. Our results reveal a strong correlation between the DM mass ($M_{W_D}$) and the reheating temperature ($T_R$) with ratio of 
around $T_R/M_{W_D} \sim 0.1$
to obtain correct DM density for detectable interaction strength. 
The decay processes are generally subdominant for the DM production but they can be important when kinematically allowed and the DM mass is close 
to half of the Higgses mass.
The DM production with DM masses below 100 GeV is driven primarily by the 
scatterings of the SM fermions and Higgses decay whereas 
the case with higher DM masses is achieved 
mainly due to the Higgses scatterings. The enhanced coupling for the strong freeze-in in our framework enables potential detection prospects
in direct and indirect detections and collider experiments. 
The parameter space of the model has already been explored partly by the current direct detection experiments and it can be explored further by future experiments such as Darwin. On the other hand, the indirect detection experiments in the current and near future are not sensitive enough to test our model.

\end{abstract}
\maketitle
\section{Introduction} \label{Intro}

There is undeniable evidence of Dark Matter (DM) in a wide range of length scales from sub-galaxies to cosmological scales.
DM accounts for over 80\% of the matter of the Universe but the origin of DM and its properties are not yet uncovered.
We have understood the existence of DM only through gravitational interactions and
the Cosmic Microwave Background observation from the Planck data indicates that the total DM relic density is given by \cite{Planck:2018vyg}
\begin{align}
\Omega_{\rm DM} h^2 &= 0.1200\pm0.001.
\end{align}
Among the variety of DM models, the Weakly Interacting Massive Particle (WIMP) is a theoretically well motivated DM candidate.
For WIMP DM, the relic density is determined by the thermal freeze-out mechanism.
To obtain the correct relic density of WIMP DM, the required total thermal averaged 
cross section is \cite{Steigman:2012nb}
\begin{align}
\langle \sigma v \rangle_{\rm tot} &= \frac{1}{ \left( 20 {\rm TeV} \right)^2}.
\end{align}
However, due to the null result of direct detection, most simple popular WIMP-type DM models have been constrained. Recent data from the direct detection experiment 
LUX-ZEPLIN \cite{LZ:2024zvo}
has put severe constraints on DM-nucleon scattering cross-section. As a result, 
the DM direct interaction 
with the SM sector is already almost ruled out if the DM annihilates into the Standard Model (SM) sector and it accounts for the total DM density.
Therefore, it is timely to look into the alternative DM production 
mechanisms which can explain 
the non-detection of DM so far as well as lead to rich future detection prospects. 

One of the most common alternative thermal DM scenarios is known as freeze-in 
mechanism \cite{Hall:2009bx}.
DM produced through the freeze-in mechanism is so-called the Feebly Interacting Massive Particle (FIMP) \cite{Hall:2009bx}.
FIMP DM assumes that DM cannot enter the thermal equilibrium in the early Universe because DM very weakly interacts with the SM particles in the early Universe. 
FIMP DM can be produced and accumulated via the energy leakage from the decay and/or scatterings of the thermal bath particles.
The relic density of FIMP DM is proportional to the thermal decay rate and/or scattering cross-section.
That is, the FIMP DM model makes a strong assumption that the pre-existing DM abundance is zero, which is questionable.
In addition, it is very difficult to test the model due to such a very small coupling 
constant. Another couple of interesting DM production includes axion dark matter which 
can be produced from the scattering of the SM bath particles \cite{Choi:2018lxt}.
Moreover, right-handed sneutrino DM can be produced by the scattering processes, due to its mixing with left-handed sneutrinos which becomes effective after electroweak symmetry 
breaks down \cite{Choi:2018vdi}. 
In the present work, we go beyond the freeze-in mechanism where we can accommodate the
stronger coupling constants by considering the low reheating temperature of the Universe.

As a matter of fact, in order for Big Bang Nucleosynthesis (BBN) to work \cite{Hannestad:2004px}, it is sufficient to take the temperature of the SM bath particles as low as 4 MeV after the reheating process is complete.
Recently, FIMP DM has been studied for the case with a low reheating temperature
in Refs. \cite{Bhattiprolu:2022sdd, Cosme:2023xpa, Cosme:2024ndc, Arcadi:2024wwg, Boddy:2024vgt, Arcadi:2024obp, Lebedev:2024vor, Lee:2024wes, Belanger:2024yoj}.
In this case, compared to the standard FIMP DM, DM is much heavier than the reheating temperature, so the thermal production of the DM   from the decays and scatterings of thermal particles would be Boltzmann-suppressed. Thus, we need sufficiently sizable couplings for FIMP DM, so there are new possibilities to test the FIMP DM models by direct detection experiments, indirect detection searches and collider searches.
Simple scenarios of DM with s-wave annihilation channels have been studied, including
Scalar DM, Fermion DM or Vector DM with the SM Higgs boson mediator \cite{Arcadi:2024wwg}  
and Fermion DM with a very light or very heavy $Z'$ mediator \cite{Boddy:2024vgt, Arcadi:2024obp}.

In the present work, we consider a renormalizable model for vector dark matter (VDM)
as the extension of the SM
with a dark $U(1)$ gauge group and a singlet scalar field. We focus on the case where the reheating temperature
of the Universe is smaller than the DM mass. In this case, the number densities of the
bath particles associated with the DM production are Boltzmann suppressed, so we need to increase the DM couplings to produce the correct DM density. In our model, we investigate both decay and scattering processes simultaneously for vector dark matter production. 
Then, we show the parameter space where the decay processes can dominate over the scattering processes, depending on the value 
of the DM mass and the reheating temperature.

We provide a detailed analysis to discuss how the decay processes of the Higgs-like scalars $h_i$ are subdominant for the strong freeze-in, given the Boltzmann suppression factors with $e^{-M_{h_{i}}/T_{R}}$ factor, except when the DM is nearly half of the masses of $h_i$. 
The scattering processes between SM fermions or Higgs-like scalars are relevant, depending on the DM mass. The low reheating temperature results in the increment in the coupling constant, making the DM detection very promising through direct, indirect and collider experiments, as compared to the standard FIMP DM. 
We constrain some parts of the parameter space by LUX-ZEPLIN and discuss how the rest of the 
parameter space will be explored in the future. 
In comparison, we also discuss the constraints and prospects of indirect detection experiments for the present model.

The paper is organized as follows. 
In Sec.~\ref{model}, we consider the UV-complete Vector Dark Matter (VDM) with a gauged $U(1)_D$ symmetry where VDM mass is generated by the spontaneous symmetry breaking with the dark Higgs.
In Sec.~\ref{Low:Heat}, we show how to achieve a low reheating temperature after inflation.
Sec.~\ref{const} deals with all the relevant constraints for present work 
including the relic density, perturbativity, collider limits, bounds from DM direct 
and indirect detections and BBN.
In Sec.~\ref{DM:results}, we present the analytic solutions including both the decay 
and scattering processes of thermal bath particles. We also present the allowed 
parameter space to satisfy the relic density of DM and DM direct detection bounds by 
randomly scanning the input parameters.   
We will conclude in Sec.~\ref{sandc}.

\section{A renormalizable model for vector dark matter}
\label{model}
We explore vector dark matter in the model with a dark $U(1)$ gauge symmetry.
The simplest realization of vector dark matter with nonzero mass can be made with 
a new SM singlet scalar, the so-called dark Higgs $\Phi$ only \cite{Lebedev:2011iq}. 
\begin{center}
\begin{table}[h!]
\begin{tabular}{||c|c|c|c||}
\hline
\hline
\begin{tabular}{c}
    Gauge\\
    Group\\ 
    \hline
    
    ${\rm SU(2)}_{\rm L}$\\  
    \hline
    ${\rm U(1)}_{\rm Y}$\\
    \hline
    ${\rm U(1)}_{\rm D}$\\      
\end{tabular}
&

\begin{tabular}{c|c|c}
    \multicolumn{3}{c}{Baryon Fields}\\ 
    \hline
    $Q_{L}^{i}=(u_{L}^{i},d_{L}^{i})^{T}$&$u_{R}^{i}$&$d_{R}^{i}$\\ 
    \hline
    $2$&$1$&$1$\\ 
    \hline
    $1/6$&$2/3$&$-1/3$\\
    \hline
    $0$&$0$&$0$\\      
\end{tabular}
&
\begin{tabular}{c|c}
    \multicolumn{2}{c}{Lepton Fields}\\
    \hline
    $L_{L}^{i}=(\nu_{L}^{i},e_{L}^{i})^{T}$ & $e_{R}^{i}$\\
    \hline
    $2$&$1$\\
    \hline
    $-1/2$&$-1$\\
    \hline
    $0$&$0$\\    
\end{tabular}
&
\begin{tabular}{c|c}
    \multicolumn{2}{c}{Scalar Field}\\
    \hline
    $H$ & $\Phi$\\
    \hline
    $2$ & $1$\\
    \hline
    $1/2$ & $0$\\
    \hline
    $0$ & $1$\\    
\end{tabular}\\
\hline
\hline
\end{tabular}
\caption{Particle contents and their corresponding
charges under the SM gauge group. }
\label{tab1}
\end{table}
\end{center}

The complete Lagrangian takes the following form \cite{Lebedev:2011iq,Ko:2014gha},
\begin{eqnarray}
\mathcal{L}& = &\mathcal{L}_{SM} -\frac{1}{4} W_{D\,\mu\nu} W^{D\,\mu\nu} 
+ (D_{\mu}\Phi)^{\dagger} (D^{\mu}\Phi) -     
  \mathcal{V}\left(H , \Phi \right)\,
\end{eqnarray}
where $W_{D\,\mu\nu} = \partial_{\mu} W_{D\,\nu} - \partial_{\nu} W_{D\,\mu}$
and the covariant derivative in the above Lagrangian for the extra 
field  is defined as 
$D_{\mu} = \partial_{\mu}  -i g_{D} n_{A} W_{D \mu}   $ where 
 $g_{D}$, $W_{D}$, $n_{A}$ are gauge coupling, gauge 
boson and charge of the field associated with the additional gauge group $U(1)_{D}$ \footnote{ For $SU(2)_D$ dark matter and  connection to $U(1)$ VDM, see Ref.~\cite{Hambye:2008bq}. }.
The potential takes the form,
\begin{eqnarray}
\mathcal{V}(H,\Phi) = - \mu^2_{\Phi} \Phi^{\dagger}  \Phi +
\lambda_{\Phi} (\Phi^{\dagger}  \Phi)^{2} - \mu^2_{H} H^{\dagger} H +\lambda_{H} (H^{\dagger} H )^{2} + \lambda_{H \Phi} (\Phi^{\dagger} \Phi) (H^{\dagger} H ).
\end{eqnarray}
When the electroweak and the $U(1)_D$ symmetry get broken, 
the scalars take the following form in the Unitary gauge,
\begin{eqnarray}
H = \frac{1}{\sqrt{2}}
\begin{pmatrix}
0 \\
v_H + h  
\end{pmatrix}\,,\,\,\,\,
\Phi =
\frac{1}{\sqrt{2}}
\begin{pmatrix}
v_{\Phi}+\phi  
\end{pmatrix}\,.
\end{eqnarray}
The mass matrix for the scalars in the basis $(h\,\,\,\phi)$ takes the following form,
\begin{eqnarray}
M_{h \phi} = \begin{pmatrix}
2 \lambda_{H} v^{2}_H & \lambda_{H \Phi} v_H v_{\Phi}\\
 \lambda_{H \Phi} v_H v_{\Phi} & 2 \lambda_{\Phi} v^2_{\Phi}
\end{pmatrix}\,.
\end{eqnarray}
Once we diagonalise the above mass matrix then we get the following mass eigenstates,
\begin{eqnarray}
\begin{pmatrix}
h_{1}\\
h_{2}
\end{pmatrix}
= \begin{pmatrix}
\cos\theta & -\sin\theta\\
\sin\theta & \cos\theta
\end{pmatrix}
\begin{pmatrix}
h\\
\phi
\end{pmatrix}.
\end{eqnarray}
The quartic couplings can be written in terms of the two Higgs masses and the mixing 
angle in the following manner,
\begin{eqnarray}
\lambda_{H} &=& \frac{M^2_{h_{1}} + M^2_{h_{2}} + \left( M^2_{h_{1}} - M^2_{h_{2}} \right) \cos 2 \alpha}{4 v^{2}_H}, \nonumber \\
\lambda_{\Phi} &=& \frac{M^2_{h_{1}} + M^2_{h_{2}} - \left( M^2_{h_{1}} - M^2_{h_{2}} \right) \cos 2 \alpha}{4 v^{2}_{\Phi}}, \nonumber \\
\lambda_{H \Phi} &=& \frac{ \left( M^2_{h_{2}} - M^2_{h_{1}} \right) \cos\alpha \sin\alpha}{v_H v_{\Phi}}.
\end{eqnarray}

After the $U(1)_{D}$
symmetry gets broken then the $U(1)_D$ gauge boson $W_{D\mu}$ acquires
the following mass,
\begin{eqnarray}
M_{W_{D}} =  g_{D} v_{\Phi}.
\end{eqnarray}
The additional gauge boson can be a suitable DM candidate once we impose a $\mathbb{Z}_{2}$ or
charge conjugation symmetry which forbids kinetic mixing 
with the $U(1)_{Y}$ gauge boson or one can consider very tiny kinetic mixing among 
the abelian gauge bosons in the model without introducing any additional 
symmetry which will make the $U(1)_{D}$ gauge
boson a stable DM candidate at the cosmological scale. 
For the stabilization of the DM, we can assume the charge conjugation symmetry 
in the dark sector which transforms the fields as follows \cite{Lebedev:2011iq},
\begin{eqnarray}
    \Phi \rightarrow \Phi^{\dagger}\,, \quad W_{D\,\mu} \rightarrow - W_{D\,\mu}\,.
\end{eqnarray}
It is very evident that the above choice of charge conjugation symmetry forbids 
the kinetic mixing between the $U(1)_D$
and the SM hypercharge gauge groups, making the additional gauge boson a stable DM candidate.

\section{Achieving Low Reheating Temperature} \label{Low:Heat}

In Refs.~\cite{Cosme:2023xpa, Belanger:2024yoj, SilvaMalpartida:2023yks}, 
the reheating temperature at a GeV scale was achieved by 
considering additional scalar as the inflaton which has a long decay lifetime. 
In Refs.~\cite{Aoki:2022dzd,Clery:2023ptm}, authors have shown
that when the inflaton is the SM Higgs doublet, it is difficult to achieve 
a very low scale reheating temperature. The general idea of achieving a low reheating
temperature can be realized when a scalar field beyond the SM \cite{Choi:2016eif} plays a role of the inflaton $\chi$ which has a long lifetime before it decays into SM particles. 
In our model, $\phi$ can be a suitable inflaton candidate and serve as $\chi$.
By following Ref. \cite{Turner:1983he,SilvaMalpartida:2023yks},
we consider the inflaton potential near the minima after the end of the inflation as follows,
\begin{eqnarray}
    V(\chi) &=& \lambda_{\chi} \frac{\chi^{n}}{\Lambda^{n-4}}\,
\end{eqnarray}
where $\lambda_{\chi}$ is a dimensionless coupling parameter, $\Lambda$ is cutoff scale
and $n$ is an integer number. After the inflation,
total energy budget of the Universe is mainly dominated by inflaton energy which
puts background equation for the inflaton and SM radiation bath to follow the 
following equations,
\begin{eqnarray}
    \frac{d \rho_{\chi}}{d a} + \frac{6 n}{n + 2} \frac{\rho_{\chi}}{a} &=& -\frac{2 n}{2+n}
    \frac{\Gamma_{\chi}}{H} \frac{\rho_{\chi}}{a},\nonumber \\
    \frac{d \rho_{rad}}{d a} + 4 \frac{\rho_{rad}}{a} &=& \frac{2 n}{2+n}
    \frac{\Gamma_{\chi}}{H} \frac{\rho_{\chi}}{a}\,.
    \label{background-eq}
\end{eqnarray}
where $a$ is the scale factor, $\rho_{\chi}$ is the inflaton energy density, $\rho_{rad}$ is the radiation energy density and  $\Gamma_{\chi}$ is the decay width of the inflaton.
The Hubble parameter $H$ is defined as
\begin{eqnarray}
    H = \sqrt{\frac{8\pi  \left( \rho_{\chi} + \rho_{rad} \right) }{3 M^2_{\rm pl}}}
\end{eqnarray}
where $M_{\rm pl} = 1.22 \times 10^{19}$ GeV.
In solving Eq.~\eqref{background-eq},
we have used the initial boundary condition $\rho_{rad} = 0$ and 
$\rho_{\chi} = \rho_{\chi}(a_{R}) \left( \frac{a}{a_{R}} \right)^{-\frac{2n}{2+n}}$ where 
$a_{R}$ is the scale factor at the reheating temperature.
At the reheating temperature $T_R$, we have $\rho_{\chi} (a_{R}) = \rho_{rad} (a_{R}) 
= \frac{\pi^2}{30}\, g_{*} \left( T_{R} \right) T_R^{4}$ where $g_{*}(T_R) \simeq 106$ is the total
relativistic {\it d.o.f} of the Universe considering the SM particle bath. 
Using the aforementioned definition and 
suitably choosing the parameters like the inflaton decay
width $\Gamma_{\chi}$, we can get the idea of the reheating temperature, $T_R$, 
by the following relation,
\begin{eqnarray}
    T_{R} = \left( \frac{\Gamma_{\chi} M_{pl}}{g^{1/2}_{*}} \sqrt{\frac{45}{8 \pi^{3}}} \right)^{1/2}.
\end{eqnarray}

\begin{figure}[h!]
\centering
\includegraphics[angle=0,height=8.5cm,width=9.0cm]{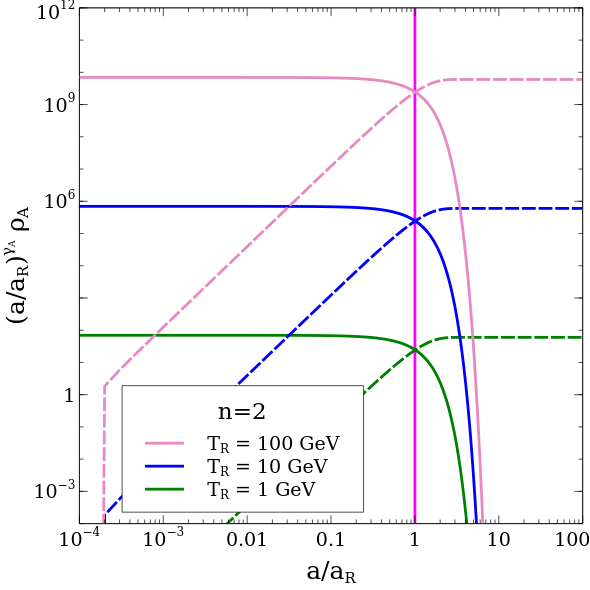}
\includegraphics[angle=0,height=8.5cm,width=8.0cm]{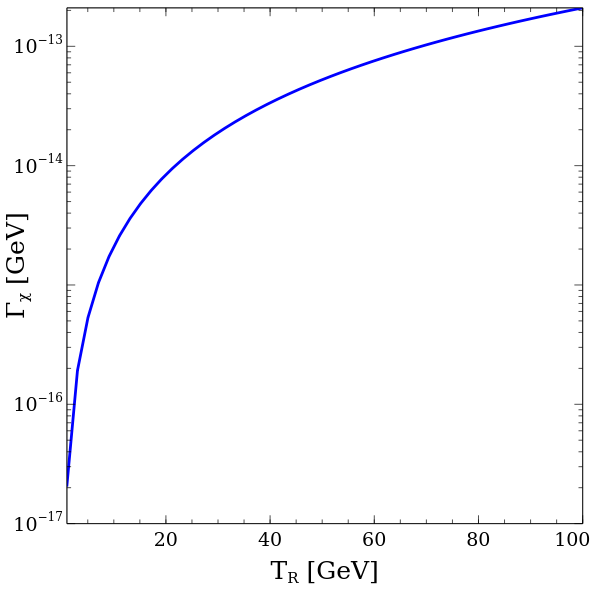}
\caption{LP shows the variation of the inflaton and radiation density with $a/a_{R}$ for a fixed value of $n=2$ whereas in the RP variation in the reheating temperature and inflaton decay width is
displayed. In the LP, A is the inflaton field $\chi$ and radiation. In the 
LP, solid lines are for the inflaton density whereas dashed lines are for SM radiation density.} 
\label{dec-reheat}
\end{figure}
In the left panel (LP) of Fig.~\ref{dec-reheat}, we present line plots for
$\frac{a}{a_{R}}$ in the x-axis and $\left( \frac{a}{a_{R}} \right)^{\gamma_{A}} \rho_{A}$
in the y-axis,
where $A = \chi, rad$ and $\gamma_{\chi} = \frac{6n}{2+n}$ while $\gamma_{rad} = 4$.
The plot is shown for $n=2$ and exhibits a similar behaviour for other values of $n$.
From the plot, we observe that at $a/a_{R} = 1$, the energy densities of both the inflaton and
radiation become equal. When the inflaton decays, the total energy density stabilizes to the total
energy content of the SM bath. Since the energy density of the thermal bath is
proportional to the fourth power of the temperature, different reheating temperatures correspond to
different amounts of radiation and inflaton energy.
In the right panel (RP) of Fig.~\ref{dec-reheat}, we illustrate the reheating temperature $T_R$ as a 
function of the inflaton decay width $\Gamma_{\chi}$. 
The plot exhibits a parabolic-like behaviour
where an increase in the decay width results in a higher reheating temperature. This implies that a
faster inflaton decay leads to an earlier reheating process with a higher reheating temperature. While
solving Eq.~\eqref{background-eq}, 
we define the reheating temperature $T_R$ as the point at which the inflaton and
radiation energy densities become equal.

\section{Constraints} \label{const}
In this section, we briefly describe the theoretical and experimental constraints relevant to our study. The main constraints arise from the perturbativity of the quartic couplings, the 
DM relic density, collider bounds from the Higgs invisible decay and Higgs signal strength, as well as DM direct and indirect detection limits and BBN bound.

\subsection{Perturbativity and bound from below conditions}
We have used the constraints on the quartic couplings, arising from perturbativity and the bounded-from-below condition of the potential which are as follows,
\begin{eqnarray}
0 \leq \lambda_{H, \Phi} \leq 4 \pi\,\quad \lambda_{H \Phi} < 4 \pi\,,\quad
\lambda_{H \Phi} \geq - 2 \sqrt{\lambda_{H} \lambda_{\Phi}}\,.
\end{eqnarray}
In our parameter space, the SM Higgs quartic coupling is always $\lambda_{H} \sim 0.13$
and the other quartic couplings are always positive, 
allowing us to easily evade all the bounds.

\subsection{DM relic density}

We have considered the DM relic density in the following range:
\begin{eqnarray}
10^{-3} \leq \Omega_{DM} h^{2} \leq 0.12.
\end{eqnarray}
This wide range has been chosen to obtain more data points while satisfying all relevant bounds 
within a reasonable time during the scanning of all the parameters. 
As we will see, any benchmark point that is underabundant in relic density can be enhanced by a slight adjustment of the reheating temperature $T_R$ and match with the relic density bound 
by Planck measurement \cite{Planck:2018vyg}. 
Additionally, varying $T_R$ will not affect the other DM observables in our study. Utilizing this dependence of DM production on $T_R$, we apply the direct or indirect detection bounds on DM by assuming 100\% vector DM, although those points could be underabundant for that particular value of $T_R$. 

\subsection{Collider bounds}
In the present model, the important bounds from the collider come from the Higgs 
signal strength and Higgs 
invisible decay. Higgs signal strength measurement implies a bound on the 
Higgs mixing angle as 
$\sin\theta < 0.23$ \cite{ATLAS:2016neq, CMS:2022dwd, Khan:2025yko}. The Higgs decay to DM is also 
severely constrained when the decay mode is kinematically open. The present-day bound on the Higgs 
invisible decay from the LHC is less than $\Gamma_{h_{1} \rightarrow inv} / \Gamma^{tot}_{h_{1}} \leq 0.16$ \cite{CMS:2018yfx, CMS:2021far, CMS:2020ulv}.

\subsection{DM Direct Detection}

\begin{figure}[h!]
\centering
\includegraphics[angle=0,height=5.0cm,width=9.0cm]{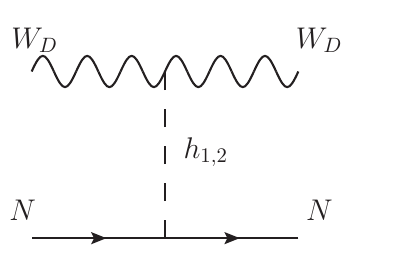}
\caption{Feynman diagram for the DM direct detection with the detector nucleus $N$
interacted by the Higgs-like scalars.} 
\label{DM-DD-fig}
\end{figure}

In this study, the VDM $W_D$ can be probed through direct detection experiments as 
shown in Fig. \ref{DM-DD-fig}. The spin-independent direct detection (SIDD) 
cross-section between DM and nucleon is given by
\begin{align}
\sigma_{\rm SI} = \frac{\mu^2 \sin^2(2\theta) g_D^2}{4 \pi v_H^2}
\left( \frac{1}{M_{h_1}^2} - \frac{1}{M_{h_2}^2} \right)^2
\left[ \frac{Z f^p_\alpha  + (A-Z) f^n_\alpha}{A} \right]^2
\,\label{dd-cs}
\end{align}
where $\mu = \frac{M_{W_D} m_N}{M_{W_D} + m_N}$ with $m_N$ denoting the nucleon mass
and $Z$ and $A$ representing the atomic and mass numbers, respectively. The proton and neutron form factors are given by,
    \begin{align}
      f^p_\alpha &= m_N \left[
      \frac{7}{9} \left(f_p^u + f_p^d + f_p^s\right) + \frac{2}{9}
      \right]
      \,,\nonumber\\
      f^n_\alpha &= m_{N} \left[
      \frac{7}{9} \left(f_n^u + f_n^d + f_n^s\right) + \frac{2}{9}
      \right]
      \,
    \end{align}
    where $f_p^u = f_n^d = 0.02$, $f_n^u = f_p^d = 0.026$, and $f_p^s = f_n^s = 0.043$ \cite{Junnarkar:2013ac}. 
The relevant SIDD constraints for the mass range considered in this study are derived 
from LUX-ZEPLIN \cite{LZ:2024zvo}. Additionally, we take into account the limits 
imposed by the proposed DARWIN detector \cite{DARWIN:2016hyl} 
and the neutrino floor \cite{Freedman:1973yd, Freedman:1977xn}.

\subsection{DM Indirect Detection}

\begin{figure}[h!]
\centering
\includegraphics[angle=0,height=4.0cm,width=14.0cm]{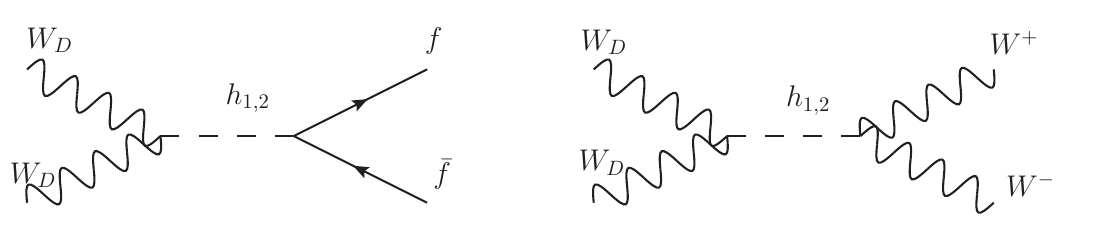}
\caption{Feynman diagrams for the DM indirect detection when it annihilates to the charged
fermions and gauge bosons mediated by the Higgs-like scalars.} 
\label{DM-ID-fig}
\end{figure}

In the present study, we have considered the indirect detection (ID) signal for
the $WW$ and $bb$ modes as shown in Fig. \ref{DM-ID-fig}. 
The non-relativistic approximation for the cross-section times
velocity can be expressed as,
\begin{eqnarray}
&& \left\langle \sigma v \right\rangle_{WW} \simeq \frac{4 M^4_{W_{D}} - 4 M^2_{W_{D}} M^2_{W}
+ 3 M^4_{W} }{96\pi M^2_{W_{D}} M^4_{W}} 
\sqrt{1 - \left(\frac{M_{W}}{M_{W_{D}}}\right)^{2}}
\biggl|\sum_{i=1,2} \frac{g_{h_{i} W_{D} W_{D}} g_{h_{i} W W}}{(4 M_{W^2_{D}} - M^2_{h_{i}}) 
+ {\bf i}\, \Gamma_{h_{i}} M_{h_{i}}} \biggr|^{2}, \nonumber \\
&& \left\langle \sigma v \right\rangle_{ff} \simeq \frac{n_{c} }{12\pi} 
\left(1 - \left(\frac{M_{f}}{M_{W_{D}}}\right)^{2}\right)^{3/2}
\biggl|\sum_{i=1,2} \frac{g_{h_{i} W_{D} W_{D}} g_{h_{i} ff}}{(4 M_{W^2_{D}} - M^2_{h_{i}}) 
+ {\bf i}\, \Gamma_{h_{i}} M_{h_{i}}} \biggr|^{2}
\label{id-cs}
\end{eqnarray}
where the vertex factors take the form,
\begin{eqnarray}
&&g_{h_{1(2) W_{D}W_{D}}} = - 2 g_{D} M_{W_{D}} \sin\theta (-\cos\theta), \nonumber \\
&&g_{h_{1(2) WW}} = \frac{e^{2} v_H }{2 \sin^{2}_{\theta_w}} \cos\theta (\sin\theta),
\nonumber \\
&&g_{h_{1(2)ff}} = \frac{M_{f}}{v_H} \cos\theta (\sin\theta),
\label{couplings}
\end{eqnarray} 
with $\sin^{2}\theta_{w} = 0.23$ and $e=\sqrt{4 \pi \alpha_{EM}}$ ($\alpha_{EM}$ being the fine 
structure constant). We have considered the future projection of 
Fermi-LAT (18Y) \cite{LSSTDarkMatterGroup:2019mwo} and CTA experiment \cite{CTA:2020qlo}. 

\subsection{BBN Bound}
In the present work, we are focusing on the low reheating temperature for DM production through the freeze-in mechanism. This provides an exponential suppression in the density of the associated particles which can be overcome by enhancing the relevant couplings. The
low reheating temperature we consider, the stronger the coupling required to achieve the correct DM density. The freedom to low reheating temperature is bounded by 
the BBN temperature which demands 
$T_{R} > 4$ MeV to preserve the successful predictions of BBN \cite{Hannestad:2004px}.
We have followed this BBN bound in our study. 

\section{Dark Matter Results} \label{DM:results}

The Boltzmann equation for producing DM from the decay and scattering processes can
be expressed as
\begin{align}
\frac{d Y_{W_{D}}}{dz} &= \frac{2 M_{\rm pl} z \sqrt{g_{*}(z)}}{1.66 M^2_{sc} g_{s}(z)}
\left[ \langle \Gamma_{h_{i} \rightarrow W_{D}W_{D}} \rangle \left( 
Y^{eq}_{h_{i}} - Y^2_{W_{D}} \right) \right] \nonumber \\
& + \frac{2 M_{sc} s(T)}{z^{2} H(T) T} \left[ Y^2_{i} \langle \sigma v \rangle_{ii \rightarrow W_{D}W_{D}} - Y^2_{W_{D}} \langle \sigma v \rangle_{W_{D}W_{D}
 \rightarrow  ii} \right] 
 \label{BE}
\end{align}
where $Y_{\alpha}=\frac{n_{\alpha}}{s(T)}$ is the co-moving number density
of species $\alpha$.
The decay and scattering contributions to the total amount of DM density 
have been displayed in Eq.~\eqref{BE}, $M_{sc}$ is any mass scale that 
can be SM or BSM Higgs mass, 
$z = \frac{M_{sc}}{T}$,
$g_{*}$ is the effective relativistic d.o.f of the Universe, and the 
local entropy is given by
\begin{eqnarray}
s(T) =  \frac{2 \pi^{2}}{45} g_{*} T^{3}.
\end{eqnarray}
The thermal average decay width and cross section times velocity are expressed as,
\begin{eqnarray}
\langle \Gamma_{h_{i} \rightarrow W_{D}W_{D}} \rangle &=& 
\Gamma_{h_{i} \rightarrow W_{D}W_{D}}\, \frac{K_{1}\left(z \right)}
{K_{2}\left(z \right)}, \\
\langle \sigma v \rangle_{ii} &=& \frac{g^2_{i} T}{2 \left( 2 \pi \right)^{4} 
n^2_{i}} \int^{\infty}_{4 \left(max[M_{i},M_{W_{D}}]\right)^2} 
\sigma_{ii \rightarrow W_{D} W_{D}} (s - 4 M^2_{i}) \sqrt{s} K_{1} 
\left(\frac{\sqrt{s}}{T} \right)\,ds   
\end{eqnarray}
where $g_{i}$ is internal {\it d.o.f} of the species $i$, 
 $\sigma_{ii \rightarrow W_{D}W_{D}}$
is the cross-section for $i i \rightarrow W_{D}W_{D}$ process where $i$ is SM particles 
and BSM Higgs. 
$K_{1}\left(z \right) \simeq \sqrt{\frac{\pi}{2 z}}\, e^{-z}$ (for $z \gg 1$) is the modified Bessel function of second kind and 
$n_{i} = g_{i} \left( \frac{M_{i} T}{2 \pi} \right)^{3/2} e^{-\frac{M_{i}}{T}}$
is the number density of species $i$ in the non-relativistic limit.
The $h_i$ ($i = 1,2$) decay rates to $W_{D}W_{D}$ can be expressed as,
\begin{eqnarray}
\Gamma_{h_{i} \to W_D W_D} = \frac{M^3_{h_{i}} g^2_{h_{i}W_{D}W_{D}}}{128 \pi M^4_{W_{D}}}
\, \sqrt{1 - \frac{4 M^2_{W_{D}}}{M^2_{h_{i}}}}\,
\left( 1 - \frac{4 M^2_{W_{D}}}{M^2_{h_{i}}} + \frac{12 M^4_{W_{D}}}{M^4_{h_{i}}} \right)
\end{eqnarray}
where the vertex factor $g_{h_{i}W_{D}W_{D}}$ is defined earlier in Eq.~\eqref{couplings}.

The total co-moving number density can be expressed as the sum 
of decay and scattering contributions in the following manner,
\begin{eqnarray}
Y_{W_{D}} = Y^{\rm Dec}_{W_{D}} + Y^{\rm Scatt}_{W_{D}}\,.
\end{eqnarray}
The decay contribution can be computed analytically as
\begin{align}
Y^{\rm Dec}_{W_{D}} &=  \sum_{i = 1,2}\frac{ 0.17 \Gamma_{h_{i}} g_{h_{i}} M_{pl}}{g^{3/2}_{*} M^2_{h_{i}}}
\biggl[ \frac{1}{4} \sqrt{  \frac{M_{h_{i}}}{T_{R}}} \left(15 
+  \frac{2 M_{h_{i}}}{T_{R}} \left( 5 +  \frac{2 M_{h_{i}}}{T_{R}} \right) \right) 
e^{- \frac{M_{h_{i}}}{T_{R}}}   + \frac{15}{8} \sqrt{\pi} 
Erfc\left( \sqrt{  \frac{M_{h_{i}}}{T_{R}}} \right)  \biggr]. 
\label{decay-contribution}
\end{align}
Here we assumed that the matter content of
the Universe starts evolving from the reheating temperature $T_R$, $\Gamma_{h_{i}}$
is the decay width of $h_i$ to DM candidate $W_D$ and $Erfc(x) \simeq \frac{2 e^{-x^{2}}}{\sqrt{\pi}
 \left(x + \sqrt{ x^{2} + \frac{4}{\pi} } \right)} $ (for $x > 0$) is the  
 complementary error function. It is worth mentioning that our analytical estimate for the 
 decay contribution coincides with the analytical expression in Ref. \cite{Hall:2009bx} for the 
 limit $T_{R} \rightarrow \infty$.
 The scattering contribution, 
 $Y^{\rm Scatt}_{W_{D}}$, involves many more
 factors and tricky to obtain the analytical estimate depending on the 
 processes are considered. We have used micrOMEGAs \cite{Alguero:2023zol} for the computation
 of scattering and decay contributions. 
 We provide the
  analytical contribution for the simple contact kind of term 
  $h_{i}h_{i} \rightarrow W_{D}W_{D}$ under some valid approximation.
  The cross-section for $\sigma_{h_{i}h_{j} \rightarrow W_{D}W_{D}}$ considering only
  the contact term is, 
\begin{align}
\sigma_{h_{i}h_{j} \rightarrow W_{D}W_{D} } &\simeq \frac{g^2_{h_{i}h_{j}W_{D}W_{D}}}{16 \pi s}
\left( \frac{s (s-4M^2_{W_{D}})}{(s - (M_{h_{i}} + M_{h_{j}} )^{2}) (s - (M_{h_{i}} - M_{h_{j}} )^{2}) } \right)^{1/2} \left( 1 + \frac{\left( s - 2 M^2_{W_D} \right)^{2}}{8 M^4_{W_{D}}} \right)
\end{align}  
where 
\begin{eqnarray}
    g_{h_{2}h_{2}W_{D}W_{D}} = 2 g^2_{D} \cos^{2}\theta\,,\,\,
    g_{h_{1}h_{1}W_{D}W_{D}} = 2 g^2_{D} \sin^{2}\theta\,,\,\,
    g_{h_{1}h_{2}W_{D}W_{D}} = -2 g^2_{D} \cos\theta \sin\theta\,.
\end{eqnarray}
The approximate expression for the co-moving number density after considering the 
$h_{2} h_{2} \rightarrow W_{D}W_{D}$ scattering with contact term 
only takes the following form,
\begin{eqnarray}
Y^{\rm Scatt}_{W_{D}} & =&  \frac{0.34 M_{pl} \cos^{4}\theta g^{4}_{D}}{(2\pi)^{4} g^{3/2}_{*}} \frac{M_{1}}{M^4_{W_{D}}}
  \biggl[ \left( 4 M^4_{1} + 3 M^4_{W_{D}} - 4 M^2_{1} M^2_{W_{D}} \right) 
  \frac{e^{-\frac{2 M_{1}}{T_{R}}}}{4 M^2_{1}} \left(1 + \frac{2 M_{1}}{T_{R}} \right) 
\nonumber \\
&& + \left( 12 M^3_{1} - 6 M_{1} M^2_{W_{D}} \right) 
\frac{e^{- \frac{2 M_{1}}{T_{R}}}}{2 M_{1}} + 15 M^2_{1} \Gamma\left(0,\frac{2 M_{1}}{T_{R}} \right) 
\label{anni-contribution}
  \biggr]
\end{eqnarray}   
where $M_{1} = {\rm max}[M_{h_{i}},M_{W_{D}}]$, $\Gamma\left(0,x \right)$ is the
incomplete gamma function and all the quantities have been defined
earlier. Finally, once we have the total comoving number density, we can use the
following relation to compute the DM relic density \cite{Edsjo:1997bg},
\begin{eqnarray}
\Omega_{W_{D}} h^{2} = 2.755 \times 10^{8} \times \left( \frac{M_{W_{D}}}{\rm GeV} \right)\times Y_{W_{D}}\,.
\end{eqnarray} 
All the results have been generated using micrOMEGAs \cite{Alguero:2023zol} 
without any approximation after writing  
our own code for micrOMEGAs demanded by the present study and before that to generate the 
CalcHEP model file \cite{Belyaev:2012qa}, we have implemented the model at the 
Feynrules package \cite{Alloul:2013bka}. We discussed our results in detail by first 
highlighting a few line plots and then scatter plots after satisfying all the
relevant constraints.  

\begin{figure}[h!]
\centering
\includegraphics[angle=0,height=8.5cm,width=8.5cm]{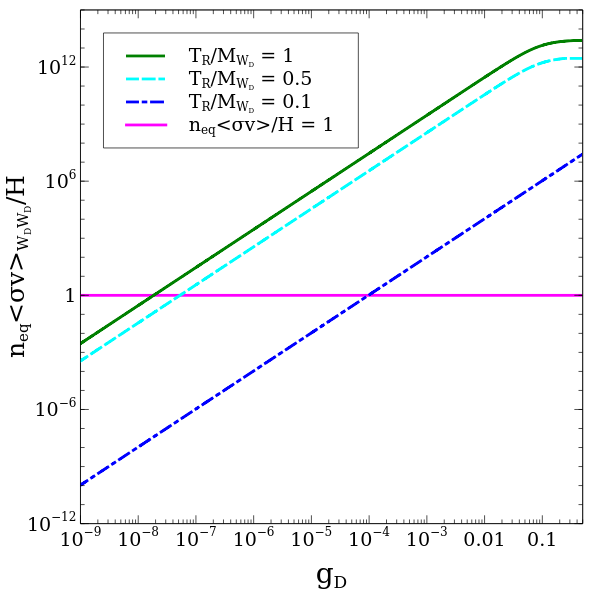}
\includegraphics[angle=0,height=8.5cm,width=8.5cm]{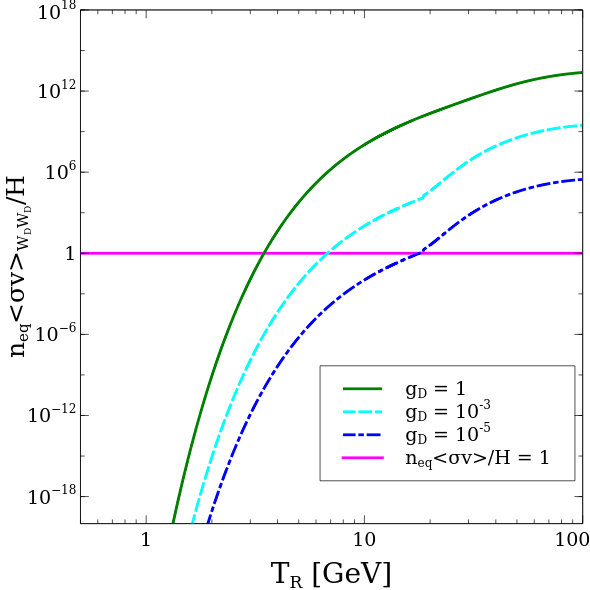}
\caption{The change of $\frac{n_{eq} \langle \sigma{}v \rangle_{W_{D}W_{D}}}{H}$ with the gauge coupling and the reheating temperature, respectively.
Additionally, each plot has been displayed for three different values of $T_{R}/M_{W_{D}}$ and $g_{D}$.
The other parameter values have been kept fixed unless they are varied,
at $M_{W_{D}} = 100$ GeV, $M_{h_{2}} = 500$ GeV, $M_{h_{1}} = 125.5$ GeV, 
$\sin\theta = 0.1$,
$g_{D} = 10^{-5}$, and $T_{R}/M_{W_{D}} = 0.1$. These
parameter values have been kept fixed for the other line plots as well.} 
\label{line-plot-1}
\end{figure}
In the left panel (LP) and right panel (RP) of Fig.~\ref{line-plot-1}, we have
shown the variation in $\left( \frac{n_{eq} \langle \sigma v \rangle_{W_{D}W_{D}}}{H} , g_{D} \right)$
and $\left( \frac{n_{eq} \langle \sigma v \rangle_{W_{D}W_{D}}}{H} , T_{R} \right)$ planes for
three different values of $T_{R}/M_{W_{D}}$ and $g_{D}$, respectively. The quantity
$n_{eq} \langle \sigma v \rangle_{W_{D}W_{D}}/H$ measures the DM
chemical equilibrium, depending on whether it is greater than $1$ or less than $1$.
In the LP, we have shown the variation with three values of $T_{R}/M_{W_{D}}$
and we notice that as it decreases, DM remains in chemical equilibrium
for higher values of the gauge coupling (green to blue line). This simply
happens because the number density $n_{eq} \propto e^{- \frac{M_{W_{D}}}{T_{R}}}$
and the cross-section times velocity
$\langle \sigma v \rangle_{W_{D}W_{D}} \propto g^2_{D} e^{- \frac{2 M_{W_{D}}}{T_{R}}} $; therefore, the reduction
in $n_{eq}$ and $\langle \sigma v \rangle_{W_{D}W_{D}}$ 
is compensated by the increase in $\langle \sigma v \rangle$ for
different values of $T_{R}/M_{W_{D}}$
to reach chemical equilibrium. 
Furthermore, when $g_D > 0.1$, a plateau appears at $T_{R}/M_{W_D}=$1 and 0.5, which results from the mutual cancellation between the Higgs-like resonance regions of $h_1$ and $h_2$.
In the RP, we have shown the variation
with $T_{R}$ for three different values of the gauge coupling $g_{D}$ 
while the DM mass is kept at
$M_{W_{D}} = 100$ GeV for both plots.
We can see that for lower values of $g_D$, higher values
of $T_R$ are needed to reach chemical equilibrium, which is equivalent to
an increase in the number density of DM. Moreover, as the reheating temperature $T_R$
approaches the DM mass, {\it i.e.}, $M_{W_{D}} = 100$ GeV,
we observe a flattening of $n_{eq} \langle \sigma v \rangle$
which implies less exponential suppression in the number density and 
$\langle \sigma v \rangle$.
Both plots indicate that increasing
$g_{D}$ allows DM to reach chemical equilibrium easily but can be avoided by
choosing the lower
values of $T_{R}/M_{W_{D}}$. This property of associating
a low reheating temperature with higher values of the gauge coupling, while avoiding
DM chemical equilibrium, allows for detection prospects of freeze-in DM.  
Additionally, a kink appears around $T_{R} \sim 20$ GeV, corresponding to the transition of dominant DM annihilation channels from $WW,~{\rm and}~ZZ$ (below) to $WW, ~h_{1}h_{1}, ~ZZ$, and $t\bar{t}$ (above).

\begin{figure}[h!]
\centering
\includegraphics[angle=0,height=8.5cm,width=8.5cm]{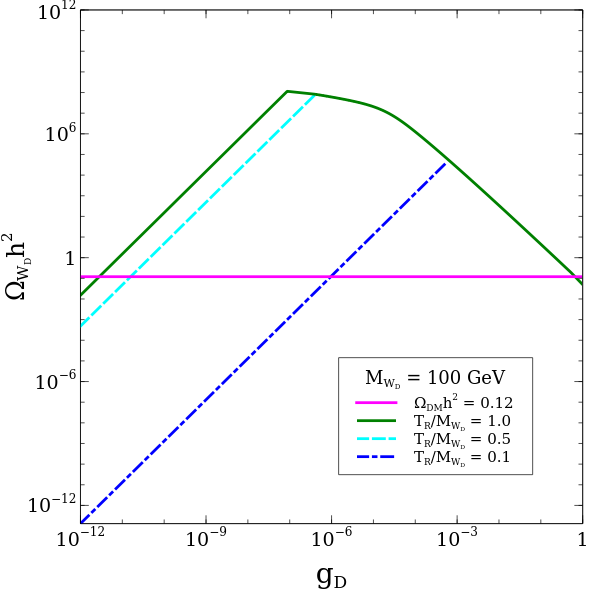}
\includegraphics[angle=0,height=8.5cm,width=8.5cm]{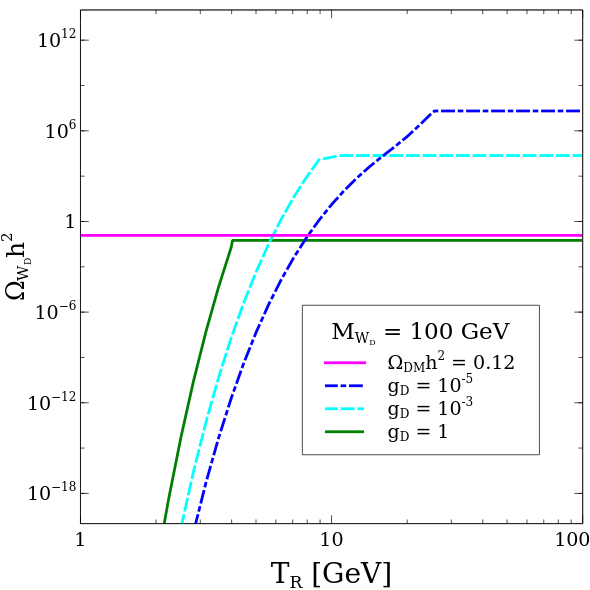}
\caption{The expected DM relic density with the gauge coupling
$g_{D}$ (Left) and the reheating temperature $T_{R}$ (Right). The three different lines in each plot represent different values of $T_{R}/M_{W_{D}}$ and $g_{D}$.
} 
\label{line-plot-2}
\end{figure}

In Fig.~\ref{line-plot-2}, we have shown the variation of
DM relic density with $g_{D}$ and $T_{R}$ for three different values of
$T_{R}/M_{W_{D}}$ and $g_{D}$ in the LP and RP, respectively. 
In both plots, we have kept the DM mass
fixed at $M_{W_{D}} = 100$ GeV as mentioned in the caption of 
Fig.~\ref{line-plot-1}.
In the LP, we can see a clear distinction
between DM production via the freeze-in and freeze-out mechanisms.
In the case of the freeze-in mechanism, the DM production rate increases with
the gauge coupling value, whereas for the freeze-out mechanism,
we expect the opposite behaviour.
To obtain the plot, we checked whether
DM reaches chemical equilibrium using our own developed code in micrOMEGAs.
After that, we solved the Boltzmann equation accordingly using
micrOMEGAs. 
Notice that we can see the change in the slopes at a certain point.
This comes from the change of DM production from freeze-in to freeze-out depending on 
$g_D$ and the $T_{R}/M_{W_D}$ values.
The important point to take from the plot is that for lower values of
$T_R$, we can obtain the correct DM density for larger values of
$g_D$. Hence, there are detection prospects for DM even if it is produced 
via the freeze-in mechanism.
In the RP, we have shown the variation of DM density with
the reheating temperature $T_R$. We can see that for different values of $g_D$, the DM
relic density varies differently. The three values of $g_D = 1, 10^{-3}$ and $10^{-5}$, 
DM reaches thermal equilibrium at different values of $T_R$ and after that
the DM relic density freezes out, as seen from the flat line. 
Once the DM reaches thermal equilibrium
its density does not change because it is independent of $T_R$
contrary to the freeze-in mechanism. The density by freeze out mechanism
changes when the interaction strength change which is fixed for each line
in the RP and we have obtained three flat lines for three different values of $g_D$.
In both plots, we have first computed the DM relic density using the freeze-out mechanism until its interaction strength is greater than the Hubble rate. We then produced the DM density using the freeze-in mechanism and extended it until it intersects with the freeze-out regime. 
It is to be noted that around the intersection point between WIMP and FIMP regimes, we need to calculate the temperature of the dark sector and compare it with the SM bath temperature. Based on the temperature comparison, we have to use the freeze-in or freeze-out mechanism for DM production, which we left for the future work. Finally, our study has been carried out when DM is completely out of equilibrium, and the intersection points between the WIMP and FIMP regimes do not affect our conclusions.

{\bf Scattered Plots}

In generating the scatter plots, we have varied the model parameters within the following ranges:
\begin{eqnarray}
&& 10^{-3} \leq \sin\theta \leq 0.2,\,\,\,\,
1\leq \left(M_{h_{2}} - M_{h_{1}} \right)[{\rm GeV}]
\leq 1000\,, \,\, \,\, 10^{-7} \leq g_{D} \leq 1, \nonumber \\
&& 1 \leq M_{W_{D}}[{\rm GeV}] \leq 1000,\,\, \,\, 10^{-4} \leq \frac{T_{R}}{M_{W_{D}}}\leq 10^{-1}\,.
\end{eqnarray}
The model parameters have been chosen in such a way that we obtain a
significant effect on DM production due to the low reheating temperature.
In particular, we have set $\frac{T_{R}}{M_{W_{D}}}$
between $10^{-4}$ and $10^{-1}$ to maximize this effect. We have observed
in the line plots that for $\frac{T_{R}}{M_{W_{D}}} = 0.5$ and $1$,
the effects of exponential suppression in number density
are not as strong as they are for $0.1$.
The other parameter choices are well studied in the
literature for both freeze-out and freeze-in DM \cite{Hall:2009bx}. In generating the
scatter plots, we have focused only on the freeze-in mechanism and not on the freeze-out
mechanism, as the latter delivers the same result as shown in the existing 
literatures \cite{Costa:2022lpy, Khan:2023uii, Khan:2024biq}.

\begin{figure}[h!]
\centering
\includegraphics[angle=0,height=8.5cm,width=8.5cm]{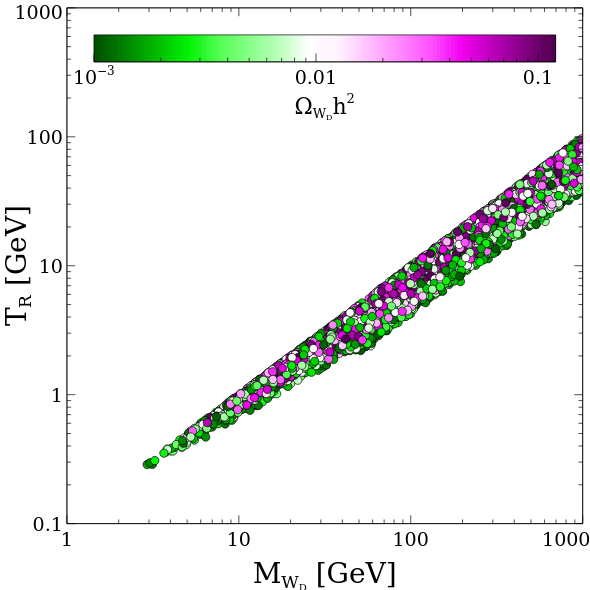}
\includegraphics[angle=0,height=8.5cm,width=8.5cm]{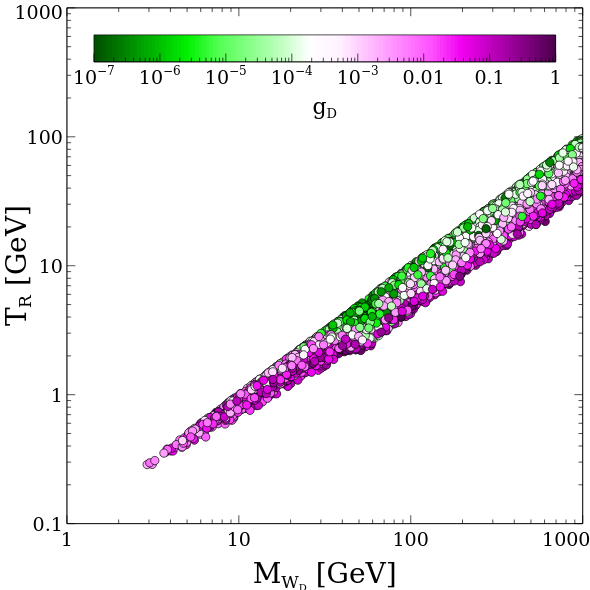}
\caption{Scattered plots in the $\left( M_{W_{D}}~,T_{R} \right)$ plane.
The color variation represents the different values of the DM relic density and the gauge coupling values. } 
\label{scatter-plot-1}
\end{figure}

In Fig.~\ref{scatter-plot-1}, we have shown the scattered plots in the
$\left( M_{W_{D}},~T_{R} \right)$ plane, where the colour variation in the LP and RP
indicates different values of DM density $\Omega_{W_{D}}h^{2}$ and 
gauge coupling $g_{D}$, respectively.
To obtain the plots, we have satisfied all the constraints discussed earlier
and imposed the requirement that the DM density lies between $10^{-3}$ and $0.12$.
We observe a linear correlation between $M_{W_{D}}$ and $T_R$ and beyond this
allowed region, DM production would exceed the considered range.
In the LP, an important aspect to note is that we do not see any colour correlation
among the points which implies that we can achieve the desired DM density
by slightly adjusting $T_R$ without significantly affecting other parameters
and observables.
Using this flexibility, when studying DM detection prospects in direct
and indirect detection later on, we do not include the fraction of DM
in the cross-section. This is because we can tune $T_{R}$ to obtain
the total DM density without altering the direct and indirect detection cross-sections.
In the RP, we observe a clear separation
between the magenta points ($g_{D} \gtrsim 10^{-3}$) and the green points
($g_{D} \lesssim 10^{-5}$). The green points appear for $M_{W_{D}} \sim 60$ GeV
and above because those points are allowed due to 
the less suppression from the exponential factor compared to the annihilation terms
as shown in Eqs. (\ref{decay-contribution}, 
\ref{anni-contribution}). The green points
for higher values of $M_{W_D} > 62.5$ GeV
mainly appears from the decay contribution of BSM Higgs $h_2$. 
\begin{figure}[h!]
\centering
\includegraphics[angle=0,height=8.5cm,width=8.5cm]{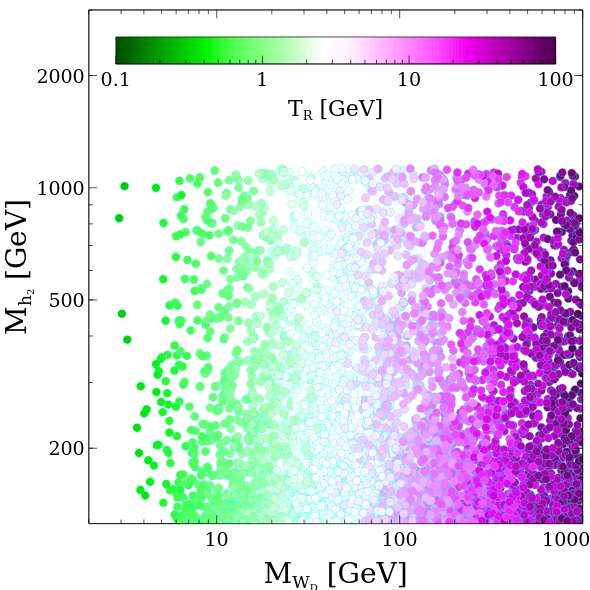}
\includegraphics[angle=0,height=8.5cm,width=8.5cm]{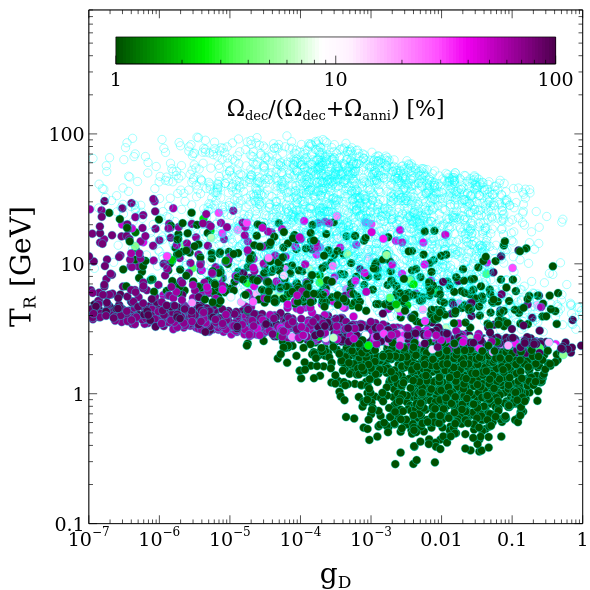}
\caption{ (Left) Scattered plot in the $\left(M_{W_{D}},~ M_{h_{2}} \right)$ plane is shown, where the colour variation indicates different values of the reheating temperature. (Right) we present the scattered plot in the $\left( g_{D},~T_{R} \right)$ plane, where
the colour variation represents the percentage contribution of Higgs decays
compared to the total decay and scattering contributions combined.} 
\label{scatter-plot-2}
\end{figure}

In Fig.~\ref{scatter-plot-2}, we present scattered plots
in the $\left(M_{W_{D}},~M_{h_2} \right)$ and $\left( g_{D},~T_{R} \right)$ planes, respectively,
where the colour variation represents different values of the reheating temperature
$T_{R}$ and the fraction of decay contribution in the DM density,
$f_{dec} = \frac{\Omega_{dec}}{\Omega_{dec} + \Omega_{anni}}$.
In the LP, we observe that all values of $M_{h_2}$ are allowed,
while $M_{W_{D}} < 2$ GeV is disallowed primarily due to the lower limit of DM 
density considered as well as the Higgs invisible decay bound.
Most of the points lie within $0.03 \leq \frac{T_{R}}{M_{W_{D}}} \leq 0.1$,
as lower values of this ratio produce a very subdominant DM component.
On the other hand, choosing a ratio greater than $0.1$ requires very low values of
the gauge coupling, aligning with the freeze-in mechanism, which is not
the focus of the present work. Due to these constraints, mainly from DM density
requirements, we find a strong correlation between the DM mass and the reheating temperature.
In the RP, we show the allowed points in the $\left( g_{D},~T_{R} \right)$ plane.
The cyan-coloured points without a colour bar represent cases where
the decay contribution is zero due to kinematic constraints.
A clear correlation emerges between $g_D$ and $T_R$ in the decay-dominated region,
which is mostly toward lower values of $g_D$. The decay production is mainly affected
by the reheating temperature by a factor proportional to $e^{-\frac{M_{h_{1,2}}}{T_R}}$,
therefore, depending on the suppression we can have dominant decay production 
for $g_{D} \sim \mathcal{O}(1)$ as well.
The green points correspond to negligible decay contributions,
indicating that scattering processes dominate.
We also observe that as $g_D$ increases, the general trend is to
lower the reheating temperature. For $g_D \sim 0.1$, there are no points with
$T_R > 40$ GeV, as such values will overproduce the DM density.
\begin{figure}[h!]
\centering
\includegraphics[angle=0,height=8.5cm,width=8.5cm]{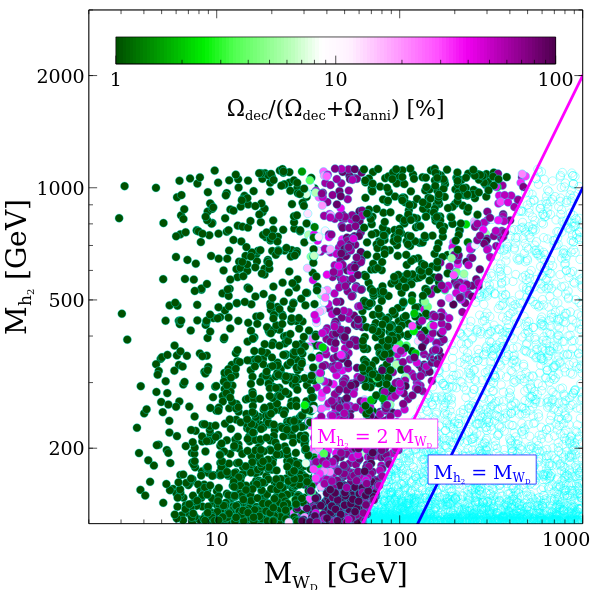}
\includegraphics[angle=0,height=8.5cm,width=8.5cm]{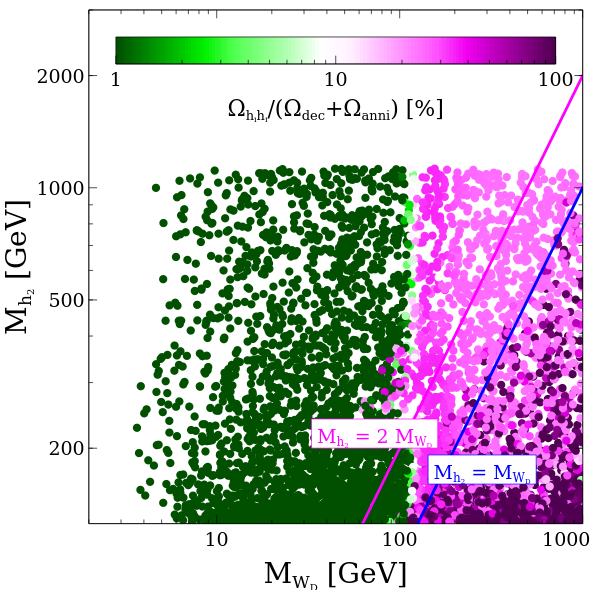}
\caption{The allowed points in the $(M_{W_{D}},~M_{h_{2}})$ region. 
In the LP, the colour variation indicates the contribution from the Higgs decay. In the RP, the plot is similar, but the colour variation represents the 
contribution from the $h_{i}h_{j}$ ($i,j=1,2$) scatterings.} 
\label{scatter-plot-3}
\end{figure}

In Fig.~\ref{scatter-plot-3}, we present scattered plots where the colour variation represents the decay contribution from the Higgs-like scalars and the scattering contribution coming solely from the Higgs-like scalars.
In the LP, the magenta points correspond to decay regions from both the SM Higgs and BSM Higgs when the decay widths are suppressed by phase space factors
as well as the exponential suppression which is less compared to the 
annihilation case.
This region also consists of a typical freeze-in scenario, requiring a relatively
high reheating temperature
and a low gauge coupling value as well as a low reheating freeze-in 
demands lower values of $T_R$ and higher values of $g_D$. 
The cyan region indicates cases where the decay width
is kinematically forbidden. The green points represent cases where the decay contribution
is subdominant and DM production is primarily driven by scattering.
The magenta and blue lines denote regions where decay is not possible and
the BSM Higgs mass is smaller than the DM mass.
In the RP, we illustrate the scattering contribution from $h_{i} h_{j}$ processes
in the colour bar.
The green points correspond to cases where this contribution is subdominant
with DM production mainly driven by the process $ff \rightarrow W_{D}W_{D}$
(where $f$ is an SM fermion) and Higgses decay. This is because 
fermionic contributions
are suppressed by a factor of $e^{-\frac{2 M_{W_{D}}}{T_{R}}}$,
whereas Higgses contributions experience even greater suppression
due to the factor $e^{-\frac{2 M_{h_{1,2}}}{T_{R}}}$,
given that the green region corresponds to $M_{h_{1,2}} > M_{W_{D}}$,
as discussed in the analytical estimates.
The magenta region which begins at $M_{W_{D}} > M_{h_{1}}$,
indicates more significant contributions from $h_{i}h_{j}$ scatterings.
In this region, we observe a $100\%$ contribution when $M_{h_{2}} < M_{W_{D}}$,
{\it i.e.}, beyond the blue line.
Therefore, we conclude from both plots that decay contributions are subdominant,
except in a very fine tuned region near the Half of the Higgses mass. 
In the other regions, scattering processes dominate
and depending on the DM mass, different annihilation modes dominate. 
The mass of DM determines the exponential suppression
factor either 
$exp[{-\frac{2\, max\left[M_{f},M_{W_{D}}\right]}{T_{R}}}]$
or $exp[{-\frac{2\, max\left[M_{h_{1,2}},M_{W_{D}}\right]}{T_{R}}}]$. 
\begin{figure}[h!]
\centering
\includegraphics[angle=0,height=8.5cm,width=8.5cm]{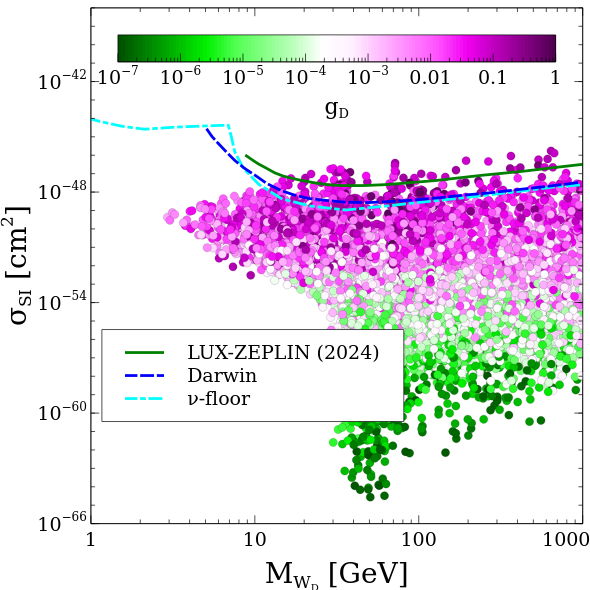}
\includegraphics[angle=0,height=8.5cm,width=8.5cm]{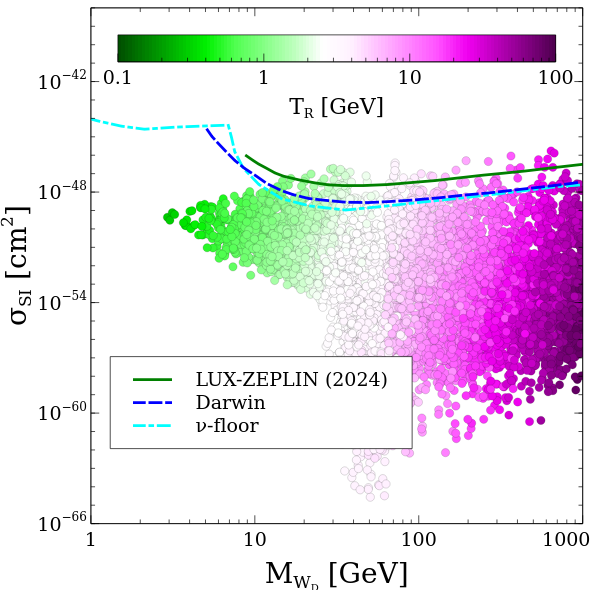}
\caption{Scattered plots in the $\left(M_{W_{D}},~\sigma_{SI} \right)$ plane, where the colour variation in the LP represents different values of the gauge coupling $g_{D}$, while in the RP, it indicates different values of the reheating temperature $T_R$.} 
\label{scatter-plot-4}
\end{figure}

In Fig.~\ref{scatter-plot-4}, we present scattered plots in the $(M_{W_{D}},~\sigma_{\rm SI})$ plane, where the colour variation in the LP corresponds to different values of the gauge coupling $g_{D}$, while in the RP, it represents the
reheating temperature $T_{R}$. The analytical expression for the SIDD 
cross section for the VDM is given in Eq.~\eqref{dd-cs}.
In the LP, we see from Eq.~\eqref{dd-cs} that the SIDD cross section is proportional to the square of the gauge coupling, which is evident from the
colour variation. An interesting aspect of the SIDD cross-section at low reheating temperatures is that only a small portion of the parameter space has been explored by LUX-ZEPLIN, leaving a large region open for future exploration—this stands in sharp contrast to present-day WIMP DM searches. In the RP, we observe that the SIDD cross section is independent of the reheating temperature, as indicated by Eq.~\eqref{dd-cs}. The vertical variation of $T_R$ with DM mass arises solely from relic density constraints.
As discussed in the LP of Fig.~\ref{scatter-plot-1}, we have considered points where the relic density falls within the range $10^{-3} \leq \Omega_{DM} h^{2} \leq 0.12$ to obtain more data points within a reasonable time frame. By slightly adjusting the reheating temperature, we can achieve the total DM density for any value in the scattered plots. Therefore, we have not multiplied by the fraction of DM and have instead considered the bound equivalent to 100\% DM. In both the plots, we have shown the recent bound from 
LUX-ZEPLIN \cite{LZ:2024zvo} (green solid line) and 
future projection from DARWIN experiment \cite{DARWIN:2016hyl} (blue dashed line) 
and $\nu-$fog (cyan dashed-dot line). 
\begin{figure}[h!]
\centering
\includegraphics[angle=0,height=8.5cm,width=8.5cm]{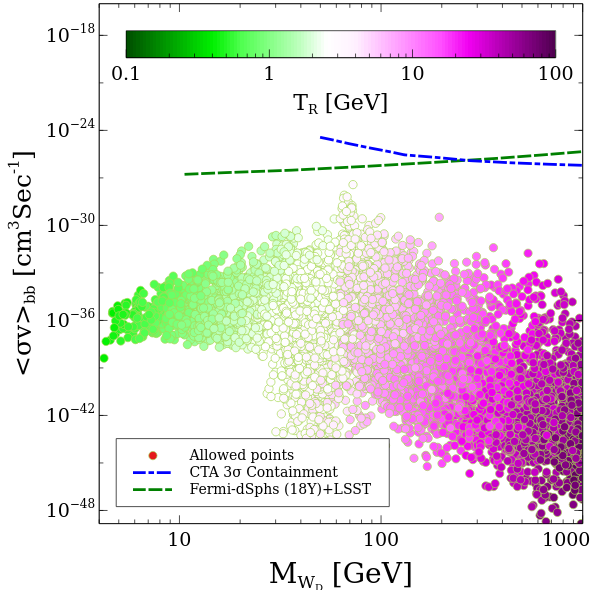}
\includegraphics[angle=0,height=8.5cm,width=8.5cm]{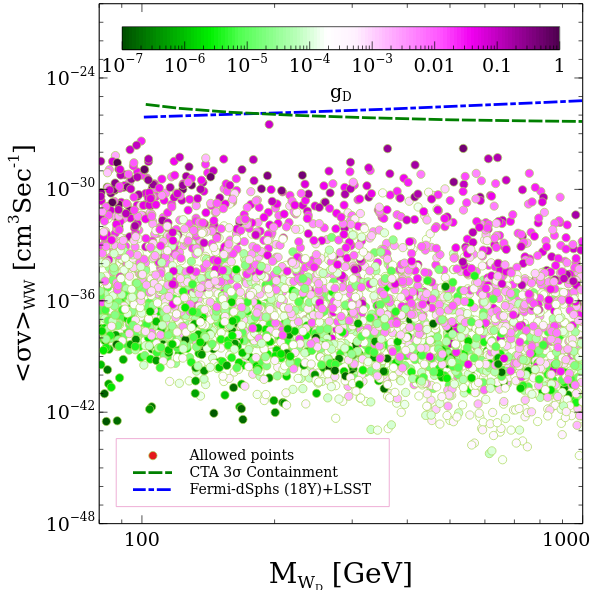}
\caption{Indirect detection signal for DM from 
$b\bar{b}$ (left) and $WW$ (right) scattering modes. The dashed line corresponds to the current DM indirect detection bounds. The colour variation in the LP corresponds to different values of $T_R$, whereas in the RP, it represents different
values of $g_D$. } 
\label{scatter-plot-5}
\end{figure}

In Fig.~\ref{scatter-plot-5}, we show the indirect detection bounds for DM. In the LP, we present results for the $b\bar{b}$ channel, while in the RP, we show results for the $WW$ channel. The analytical expressions for VDM annihilation to $b\bar{b}$ and $WW$
channels are given in Eq.~\eqref{id-cs}. 
The colour variation in the LP represents different values of the reheating temperature $T_R$, whereas in the RP, it corresponds to different values of the gauge coupling $g_D$. In the LP, we observe that the indirect detection cross-section does not depend on 
the reheating temperature, as it only affects the DM relic density, leading to a vertical correlation.
In the RP, the colour variation highlights different values of $g_D$, and we see that the cross-section increases with the gauge coupling. In both plots, we include future 
prospects for the CTA and Fermi-dSphs data, which are close to probing the present model parameter space. We anticipate that future, more sensitive indirect detection experiments or upgraded versions of the proposed experiments will further explore this model.

\section{Summary and Conclusion}
\label{sandc}
In this work, we have considered vector DM production in standard cosmology at a low reheating temperature. We have briefly discussed how  to achieve a low reheating temperature by a long inflaton lifetime with its prolonged decays into the SM particles. 
Our primary focus has been on making the coupling strength  of FIMP DM
detectable in ongoing DM experiments.
As compared to the standard freeze-in DM production which requires a very feeble coupling 
for DM, we achieve freeze-in DM production with stronger coupling at a 
low reheating temperature. In this setup, the reduction in the number density due to 
the low reheating temperature (compared to the associated mass scale) plays a crucial role. 
The thermal averages for the freeze-in DM production rates with bath particles contain the Bessel function $K_{1}\left(\sqrt{s}/T \right)$, where $s$ varies
from the squared sum of either initial or final state particle masses, depending on 
which one is greater. 
The suppression caused by the Bessel function can be compensated by the increasing in the DM coupling strength. Consequently, we can achieve the freeze-in production for vector dark matter at a strong coupling.

We have varied the model parameters to study vector DM production in detail. To this, we have taken into account all the relevant constraints, including those from DM relic density, collider bounds, BBN bound and various DM detection experiments. We have presented a few line plots demonstrating how the DM production varies with changes in the extra gauge coupling and the reheating temperature. Additionally, we have examined the DM thermalization depending on the reheating temperature and its effect on the DM production.
We have also explored the correlations among the model parameters and found a strong relationship between 
the DM mass and the reheating temperature. This correlation implies that the reheating temperature must 
lie within a specific range compared to the DM mass for the correct relic density
and detectable interaction strength. 
Furthermore, our findings suggest that the optimal ratio 
between the DM mass and the reheating temperature is approximately $T_{R}/M_{W_D} \sim 0.1$ for the detectable coupling strength.

In most of the parameter space, we found that the DM production from the decay processes is highly subdominant. However, in a specific region, DM production occurs through decay when the DM mass is around half of the Higgses mass. 
Beyond the resonance region DM production from the decay is suppressed by the 
factor $e^{-M_{h_{1,2}}/T_{R}}$ and annihilation dominates.
 Within the considered model parameter range, we observed that the production of DM with mass below 100 GeV is induced primarily by the scattering of the SM fermions
 and decay, whereas the production of heavier DM can occur via the scatterings 
 of Higgs-like scalars, depending on the relative masses of the Higgs-like scalars and DM.

Finally, due to the increased coupling for DM, we showed that DM particles can 
be detected in various 
direct and indirect detection experiments. In our analysis, we have not multiplied 
the direct and indirect detection signals by the fraction of the DM density to the observed one because a small adjustment in the reheating temperature can easily make the DM 
density to the correct value without affecting the direct and indirect detection cross-sections. 
Despite being produced via freeze in mechanism, FIMP DM can already be 
excluded by DM direct detection experiments in some parts of the parameter space. 
Moreover, we have examined the prospects for direct detection, highlighting the potential reach of the DARWIN experiment and the neutrino fog.  We have also explored indirect detection bounds, which are not important at present but may be meaningful in the future with improved sensitivity. Furthermore, we can explore further the parameter space for strong freeze-in at LHC or proposed FCC/ILC/CLIC 
colliders in the future \cite{Acar:2016rde, Golling:2016gvc, ILC:2013jhg, CLICDetector:2013tfe, Suarez:2022pcn, Agapov:2022bhm}.

\section*{Acknowledgements}
The work is supported in part by Basic Science Research Program through the National Research Foundation of Korea (NRF) funded by the Ministry of Education, Science and Technology (NRF-2022R1A2C2003567 and RS-2024-00341419 (JK)).
The work is also supported in part by Brain Pool program funded by the Ministry of Science and ICT through the National Research Foundation of Korea (RS-2024-00407977) (SK).

\bibliographystyle{utphys}
\bibliography{ref}

%

\end{document}